\documentstyle[preprint,aps]{revtex}

\begin{document}
\draft
\preprint{}
\title{How to determine approximate Mixmaster parameters\\ from
numerical evolution of Einstein's equations}
\author{Beverly K. Berger}
\address{Physics Department, Oakland University, Rochester, Michigan
48309}
\date{\today}
\maketitle
\begin{abstract}
To assess the validity of the Belinskii, Khalatnikov, and Lifshitz (BKL)
approximation to Mixmaster dynamics, it would be useful to evaluate the
BKL discrete parameters as a byproduct of the numerical solution of
Einstein's equations.  An algorithm to do this and results for a typical
trajectory are presented.
\end{abstract}
\pacs{98.80.Dr, 04.20.Jb}

\narrowtext

It has long been known that the approach to the singularity of solutions to
Einstein's equations for (e.g.) vacuum, diagonal, Bianchi Type IX
(Mixmaster) cosmologies can be described with high accuracy as an
infinite sequence of Kasner epochs \cite{BKL}.  The change in Kasner
epoch can be understood as a bounce of the trajectory off the
minisuperspace (MSS) potential \cite{Misner}.  In analogy to the
Poincar\'e map, the Belinskii, Khalatnikov, and Lifshitz (BKL) map
\cite{BKL} replaces the exact trajectory with a discrete sequence of
bounce points.  Recent interest in Mixmaster
dynamics has centered on an apparent conflict between the known chaotic
properties of the BKL map \cite{Barrow,Sinai} and vanishing
Liapunov exponents (LE's) obtained from numerical solutions to the
Mixmaster Einstein
equations \cite{Matsas,Burd,Hobill}.  Although this discrepancy is now
understood to arise from the dependence of the LE on choice of time
variable \cite{Berger1,Ferraz,Sinai}, the details of the BKL map as an
approximation to the actual solution have not yet been fully analyzed.  As
a tool toward this end, it would be helpful to generate the BKL map
parameters (with high precision) during the numerical evolution of
Einstein's equations.  This has already been done for the BKL variable $u$
which parametrizes the Kasner indices \cite{Rugh,Berger1}, but has not
yet been achieved for the other BKL variables.

Here we shall give a prescription to obtain the complete 4-parameter
BKL map
\begin{equation}
(u_{n+1},v_{n+1},p_{\Omega,n+1},\Omega_{S,n+1})=
f(u_{n},v_{n},p_{\Omega,n},\Omega_{S,n})
\end{equation}
from the numerical evolution.  (The parameters will be defined below.)
If the numerical evolution can be shown to be
highly accurate (e.g.~from accurate preservation of the Hamiltonian
constraints), deviations of the measured BKL map parameters from the
predicted ones can be used to analyze the validity of the approximations
made in the map's derivation.

The Mixmaster cosmology is described by three scale factors $a$, $b$,
and $c$ or, equivalently, by the logarithmic volume $\Omega$ and
anisotropies $\beta_{\pm}$ \cite{BKL,Misner}.  The two variable
choices are related by
\begin{mathletters}
\label{transformation}
\begin{equation}
\alpha =\ln a=\Omega -2\beta _+,
\end{equation}
\begin{equation}
\zeta =\ln b=\Omega +\beta _++\sqrt 3\beta _-,
\end{equation}
\begin{equation}
\gamma =\ln c=\Omega +\beta _+-\sqrt 3\beta _-.
\end{equation}
\end{mathletters}
Einstein's equations in the BKL time $\tau$ defined by lapse
$N=e^{3\Omega}$ are obtained by variation of
\begin{eqnarray}
H=-p_\Omega ^2+p_+^2+p_-^2+\xi\, {\cal U}(\Omega ,\beta _+,\beta _-)
\label{H}
\end{eqnarray}
where
\begin{eqnarray}
\cal U&=&e^{4\Omega }\left[ {e^{-8\beta _+}+\,2\;e^{4\beta _+}
(\cosh 4\sqrt 3\beta _--1)} \right. \nonumber \\
&&\;\left. {-\;4\;e^{-2\beta _+}\cosh 2\sqrt 3\beta _-
} \right]\nonumber \\
  &=&e^{4\alpha }+e^{4\zeta }+e^{4\gamma }-
2e^{2(\alpha +\zeta )} \nonumber \\
&&-
2e^{2(\zeta +\gamma )}-2e^{2(\gamma +\alpha )
\label{potential}
\end{eqnarray}
for $p_\Omega$, $p_{\pm}$  respectively conjugate to $\Omega$,
$\beta_{\pm}$ and $\xi$  an arbitrary constant.  If $\cal U$ is negligible,
the solution is approximately the Kasner universe with scale factors $a_i =
t^{2p_i}$ for $i=1,\,2,\,3$ and $\sum\nolimits_{i=1}^3
{p_i}=1=\sum\nolimits_{i=1}^3 {p_i^2}$.  The factor $e^{4\Omega}$ in
(\ref{potential}) approaches zero as $\Omega \to - \infty$ (the singularity)
so that $\cal U$ is negligible unless one of $\alpha$, $\zeta$, or $\gamma$
vanishes (or nearly does so).  If $\cal U$ is approximated by its dominant
term, the scattering problem may be solved exactly \cite{BKL}.  This
yields sufficient information to relate the two asymptotic Kasner
universes.

BKL found \cite{BKL} that if the Kasner indices are parametrized by
$u\in [1,\infty )$ through
\begin{mathletters}
\label{pdefine}
\begin{equation}
p_1=-u/(u^2+u+1),
\end{equation}
\begin{equation}
p_2=(u+1)/(u^2+u+1),
\end{equation}
\begin{equation}
p_3=u(u+1)/(u^2+u+1),
\end{equation}
\end{mathletters}
and the $n^{th}$ Kasner epoch is labeled by $u_n$,
then
\begin{eqnarray}
u_{n+1}=\left\{ \matrix{u_n-1\quad,\quad\quad2\le u_n<\infty \hfill\cr
  {1 \over {u_n-1}}\quad,\quad\quad1\le u_n\le 2\hfill\cr} \right.
\label{umap}
\end{eqnarray}
parametrizes the new epoch.  The rule for $1 \le u \le 2$ denotes an era
change and is the source of sensitivity to initial conditions in the evolution
toward the singularity.  It is also shown that the relationship between
$\Omega$ and $\tau$ defined by $
{{d\Omega } / {d\tau }}=-p_\Omega$
(obtained by the variation of (\ref{H})) yields
\begin{eqnarray}
p_{\Omega ,n+1}=p_{\Omega ,n}\left( {{{u_n^2-u_n+1} \over
{u_n^2+u_n+1}}} \right).
\label{pwmap}
\end{eqnarray}
In the BKL notation, $p_\Omega = - \Lambda/3$ so that (\ref{pwmap}) is
derived from the BKL relation
$\Lambda_{n+1}=\Lambda_n\,\left(2\,p_1+1\right)$ with $p_1$ expressed
in terms of $u$ through (\ref{pdefine}).
To proceed further, BKL assume the following \cite{BKL,Sinai}:

(a)  Between bounces, the evolution is exactly the Kasner solution;

(b)  The bounces occur at a fixed value of $\Omega$ (i.e.~they are
instantaneous in $\Omega$);

(c)  At a bounce, one of $\alpha$, $\zeta$, or $\gamma$ equals zero;

(d)  The MSS potential is replaced by
\begin{equation}
{\cal U} = e^{4\,\alpha}+e^{4\,\zeta}+e^{4\,\gamma}.
\end{equation}

In the following, we shall use the version of the BKL map given in
\cite{Berger2} (with some modifications).  This is itself based on Chernoff
and Barrow's version \cite{CB} of BKL's original map \cite{BKL,Sinai}.
We assume that a bounce has occurred at $\tau_0$, labeled by $\Omega_
0$, where $\zeta_0=0$.  The logarithmic scale factors (LSF's) $\alpha$,
$\zeta$,
and $\gamma$ are identified for each epoch as (evolving toward the
singularity) the increasing, decreasing, and faster decreasing components.
Parameters  $\bar u$ and $\bar v$ are defined so that
\begin{mathletters}
\label{firstbounce}
\begin{equation}
\alpha _0={{3\Omega _0} \over {1+\bar u\,(\bar v-1)}}
\end{equation}
and
\begin{equation}
\gamma _0={{3\Omega _0\,\bar u\,(\bar v-1)} \over {1+\bar u\,(\bar v-
1)}}
\end{equation}
\end{mathletters}
satisfy the identity (see (\ref{transformation}))
$\alpha + \zeta + \gamma = 3\,\Omega$
valid for all $\tau$ by definition.  The Kasner indices (\ref{pdefine})
describe the evolution of the LSF's:
\begin{mathletters}
\label{evolve}
\begin{equation}
\alpha(\tau)=\alpha_0+3\,p_1\,(\Omega-\Omega_0),
\end{equation}
\begin{equation}
\zeta(\tau)=3\,p_2\,(\Omega-\Omega_0),
\end{equation}
\begin{equation}
\gamma(\tau)=\gamma_0+3\,p_3\,(\Omega-\Omega_0)
\end{equation}
\end{mathletters}
where we have used the BKL assumptions and the fact that
$\Omega -\Omega _0=({\Lambda / 3})(\tau -\tau _0)$.
(Note that the factor of 3 in (\ref{evolve}) was inadvertently omitted in
\cite{Berger2}.)  Evolving until $\alpha_1=0$ at $\Omega_1$ yields
\begin{mathletters}
\label{secondbounce}
\begin{equation}
\zeta _1={{3\Omega _0\,\bar u\,} \over {(\bar u-1)\kern 1pt\{1+\bar u\kern
1pt(\bar v-1)\}}},
\end{equation}
\begin{equation}
\gamma _1={{3\Omega _0\,\bar u\,\bar v} \over {1+\bar u\kern 1pt(\bar v-
1)}}
\end{equation}
\end{mathletters}
with
\begin{equation}
\Omega _1=\Omega _0+{{3\Omega _0\,(\bar u^2-\bar u+1)} \over {(\bar
u-1)\,\{1+\bar u\kern 1pt(\bar v-1)\}}}.
\label{wmap}
\end{equation}
Equations (\ref{firstbounce}) and (\ref{secondbounce}) can be inverted to
obtain
\begin{mathletters}
\label{uvbar}
\begin{equation}
\bar u={{\zeta _1} \over {\zeta _1-\alpha _0}},
\end{equation}
\begin{equation}
\bar v={{\gamma _1} \over {\gamma _1-\gamma _0}}.
\end{equation}
\end{mathletters}
If, for later convenience, we define
\begin{mathletters}
\label{uv}
\begin{equation}
u = \bar u - 1,
\end{equation}
\begin{equation}
v = \bar v + 1
\end{equation}
\end{mathletters}
then $u$ is evolved by the map (\ref{umap}) while (in the presence of the
map for $u$)
\begin{eqnarray}
v_{n+1}=\left\{ \matrix{\;v_n+1\;,\quad\quad2\le u_n<\infty \hfill\cr
  {1 \over {v_n}}+1\;\;,\quad\quad1\le u_n\le 2\hfill\cr} \right..
\label{vmap}
\end{eqnarray}
We note that the map for $v$ is the inverse of that for $u$ so that in the
expansion direction $1 \le v \le 2$ corresponds to an era change with
sensitivity to initial conditions.  This has been discussed in detail
elsewhere \cite{Berger2}.  The rule for evolution of
$\Omega_S$, the value of $\Omega$ at the initial bounce of each epoch, is
just Eq.~(\ref{wmap}).

It is also useful to consider the era scale parameter given by
\begin{equation}
\kappa=\gamma_1-\gamma_0
\label{kdefine}
\end{equation}
which was shown (within the BKL approximation) to be constant over all
epochs of a given era \cite{Berger2}.  It evolves according to
\begin{equation}
\kappa_{N+1}=\bar u_{N+1}\, \bar v_N \,\kappa_N
\label{kevolve}
\end{equation}
where $N$ labels the era.  (In contrast to \cite{Berger2}, we have assigned
to the $N^{th}$ era, the value of $v$ associated with the last epoch of that
era.)  In \cite{Berger2}, it was shown how the ratio of bounce values of
$\alpha$, $\zeta$, or $\gamma$ to $\kappa$ determines the era change.

In \cite{Berger2}, the Kasner parameters $\bar u$, $\bar v$, and $\kappa$
were found from existing numerical data using (\ref{uvbar}) and
(\ref{kdefine}).  The bounce point values were computed as the intersection
of successive Kasner trajectories.  This procedure naturally gave the
sequence for $\bar u$ and $\bar v$.  The era changes and identifications of
LSF's as $\alpha$, $\zeta$, or $\gamma$ (based on
relative rate of increase or decrease) was done by hand.  The following
algorithm yields the Kasner parameters $u$, $v$, $\Omega_S$, and
$\kappa$ as a byproduct of the numerical evolution of Einstein's
equations.  Since the equations used are those obtained by variation of
(\ref{H}), the remaining BKL parameter $p_\Omega$ is found
automatically.

1.  At any point on the trajectory, $\alpha$, $\zeta$, and $\gamma$ are
computed from $\Omega$ and $\beta_{\pm}$ using (\ref{transformation}).

2.  During an approximate Kasner era, $p_\Omega$ and $p_{\pm}$ will
be (nearly) constant.  Define
\begin{mathletters}
\label{sine}
\begin{equation}
\sin \theta ={{p_-} \over {p_\Omega }},
\end{equation}
\begin{equation}
\cos \theta ={{p_+} \over {p_\Omega }}.
\end{equation}
\end{mathletters}
(These will satisfy $\sin^2 \theta + \cos^2 \theta = 1$ only to the extent
that the Kasner approximation is valid.) Then (\ref{transformation}) implies
 \begin{mathletters}
\label{ptotheta}
\begin{equation}
p_1={1 \over 3}\left( {1+2\cos \theta } \right),
\end{equation}
\begin{equation}
p_2={1 \over 3}\left( {1-\cos \theta -\sqrt 3\sin \theta } \right),
\end{equation}
\begin{equation}
p_3={1 \over 3}\left( {1-\cos \theta +\sqrt 3\sin \theta } \right).
\end{equation}
\end{mathletters}
Thus the Kasner indices are computed at all points of the trajectory.

3.  Order the $p_i$'s from smallest to largest.  Call the LSF's
associated with this ordering
$\bar \alpha$, $\bar \zeta$, and $\bar \gamma$ respectively---if, for
a given epoch, $p_2 < p_1 < p_3$ then, for that epoch, $\bar \alpha =
\zeta$, $\bar \zeta = \alpha$, and $\bar \gamma = \gamma$.

4.  Eq.~(\ref{uvbar}) for $\bar u$ and $\bar v$ can be expressed through the
evolution equations (\ref{evolve}) in terms of quantities evaluated at every
point of the trajectory.  For example, if $p_1 < p_2 < p_3$,
\begin{equation}
u=- 1 - {p_3 \over p_1},
\label{ueq}
\end{equation}
\begin{equation}
v={p_2 \over p_3}\,{{ \left(p_1 \bar \gamma - p_3 \bar \alpha\right)}
\over {\left(p_1 \bar \zeta - p_2 \bar \alpha \right)}} + 1,
\label{veq}
\end{equation}
\begin{equation}
\Omega_0 = \Omega - {\bar \zeta \over {3\, p_2}},
\label{w0eq}
\end{equation}
\begin{equation}
\Omega_1 = \Omega - {\bar \alpha \over {3\, p_1}},
\label{w1eq}
\end{equation}
and
\begin{equation}
\kappa = p_3\, \left( {\bar \zeta \over p_2} - {\bar \alpha \over p_1}
\right).
\label{keq}
\end{equation}
We again emphasize that the quantities on the right hand sides of
Eqs.~(\ref{ueq})--(\ref{keq}) are evaluated automatically at each point of
the trajectory obtained by the evolution of Einstein's equations.  Near the
maximum of expansion and during the bounces, when the BKL
parameters are invalid or meaningless, nonsensical results are obtained.
However, between bounces, when the evolution is nearly the Kasner
solution, all the parameters found from (\ref{ueq})--(\ref{keq}) and
$p_\Omega$ are (very nearly) constant.  It is then possible to compare the
measured parameter values to the predicted ones in order to study the BKL
map's validity.

Data obtained with this algorithm are shown in the Tables.  The
sequences for $u$ and $p_\Omega$ agree with predicted values (from
(\ref{umap}) and (\ref{pwmap}) respectively) to higher accuracy than do
those for $v$ (from (\ref{vmap})) and $\Omega_0$ (where a given epoch's
$\Omega_1$ should correspond to the next epoch's $\Omega_0$).  This is
because the derivation of the evolution of $u$ and $p_\Omega$ requires
only the BKL assumption (a) that the interbounce evolution be Kasner.  To
evolve $v$, $\kappa$, and $\Omega_S$ requires all the BKL assumptions
about the details of the bounce.  (Note that the algorithm produces an
incorrect value for $\Omega_0$ at the start of a new era.  This is a
consequence of incorrect matching of scale factors which interchange
roles here.  This need not be corrected since $\Omega_1$ of the previous
epoch is the desired value.)  As the singularity is approached, the true
bounce becomes closer to the BKL idealization since the MSS potential
$\cal U$ becomes smaller and smaller except at the (now more precisely
localized) bounces \cite{Sinai}.  For comparison, an approximate value of
$\Omega$ at the bounce was determined from the data.  (A more precise
value could be obtained with additional effort by reporting more points of
the trajectory.)

We note that the discrepancies in the Tables can arise both from
numerical error and from failure of the BKL map to be completely valid.
The evidence from the data is that both the difference
between $\{u,\,p_\Omega\}$ and $\{v,\,\kappa,\,\Omega_S\}$ and the
improvement as $\Omega \to - \infty$ reflect increasing validity of the
BKL assumptions.  Now that an algorithm exists to compute the BKL
parameters, precise numerical comparison between the map evolution and
the bounce details assumed in the approximation has become possible.

\acknowledgments

This work is supported in part by NSF Grant PHY-930559 to Oakland
University.  The author is grateful for hospitality the Institute for
Geophysics and Planetary Physics at Lawrence Livermore National
Laboratory.  Computations were performed with the facilities of the
National Center for Supercomputing Applications at the University of
Illinois.

\widetext
\begin{table}
\caption{Evaluated and predicted BKL parameters:  $u$, $v$, and
$\kappa$ are evaluated directly from the numerical trajectories using
(\protect \ref{ueq}), (\protect \ref{veq}), and (\protect \ref{keq}).
The predictions are based on the maps
(\protect \ref{umap}) and (\protect \ref{vmap}) for $u$ and $v$
respectively and on
(\protect \ref{kevolve}) for the update of $\kappa$.
Values of $u$ and $v$ indicated
by the same symbol should be multiplied together to evolve
$\kappa$.  (Predicted values for adding and subtracting 1 are not shown.)}
\label{table1}
\begin{tabular}{cdddddd}
epoch&$u$&$u_{\rm pred}$&$v$&$v_{\rm
pred}$&$\kappa$&$\kappa_{\rm pred}$\\
\tableline
a&7.229326& &2.03968& &$-$68.250& \\
b&6.231434& &3.11097& &$-$69.036& \\
c&5.231434& &4.14907& &$-$69.680& \\
d&4.231435& &5.16301& &$-$70.150& \\
e&3.231436& &6.16326& &$-$70.407& \\
f&2.231437& &7.16359& &$-$70.395& \\
g&1.231441& &8.18506*& &$-$70.012& \\
h&4.320915*&4.3208&1.12093&1.1222&$-$2466.6&$-$
2476.0\\
i&3.320981& &2.11641& &$-$2464.2& \\
j&2.320982& &3.11380& &$-$2461.2& \\
k&1.320984& &4.11352\dag& &$-$2457.9& \\
l&3.115458\dag&3.1153&1.24294&1.2431&$-$31472.&$-$
31499.\\
m&2.115468& &2.24244& &$-$31468.& \\
n&1.115469& &3.24229\ddag& &$-$31462.& \\
o&8.660378\ddag&8.6603&1.30840&1.3084&$-$883330.&$-$
883610.\\
\end{tabular}
\end{table}

\begin{table}
\caption{Additional BKL parameters.  $\Omega_0$ and $\Omega_1$
represent the initial and final $\Omega$ values of the indicated Kasner
epoch computed numerically using (\protect \ref{w0eq}) and
(\protect \ref{w1eq}).  The
predicted evolution is that $\Omega_1$ for the $n^{th}$ epoch should
equal $\Omega_0$ for the $n+1^{st}$ epoch.
$\Omega_{1\,{\rm meas}}$ is the
$\Omega$ value in the data closest to the bounce.  $p_\Omega$ is reported
as one of the dependent variables in the ODE solver with the predicted
value calculated from (\protect \ref{pwmap}).}
\label{table2}
\begin{tabular}{cddddd}
epoch&$\Omega_0$&$\Omega_1$&$\Omega_{1\,{\rm meas}}$&$p_\Omega$&$p_{\Omega \,
{\rm pred}}$\\
\tableline
a&$-$3.6671&$-$26.798&$-$27.603&5.3404& \\
b&$-$28.748&$-$52.270&$-$53.545&4.0646&4.0640\\
c&$-$53.643&$-$77.582&$-$78.895&2.9649&2.9649\\
d&$-$78.431&$-$102.87&$-$103.32&2.0416&2.0416\\
e&$-$103.25&$-$128.44&$-$127.60&1.2948&1.2948\\
f&$-$128.43&$-$155.14&$-$155.21&0.72453&0.72451\\
g&$-$154.80&$-$186.63&$-$186.23&0.33072&0.33072\\
h&568.25&$-$289.72&$-$287.16&0.11339&0.11339\\
i&$-$285.70&$-$1164.4&$-$1168.2&0.072547&0.072546\\
j&$-$1160.8&$-$2087.6&$-$2105.2&0.041155&0.041157\\
k&$-$2084.6&$-$3171.2&$-$3165.5&0.019216&0.019216\\
l&5393.0&$-$5915.9&$-$5922.8&0.0067300&0.0067300\\
m&$-$5909.9&$-$17991.&$-$17988.&0.0036961&0.0036961\\
n&$-$17986.&$-$32918.&$-$32925.&0.0016359&0.0016360\\
o&173160.&$-$124810.& &0.00054963&0.00054963\\
\end{tabular}
\end{table}

\end{document}